\documentclass[12pt]{iopart}

\usepackage{graphicx} 
\usepackage{bbm}
\usepackage{amssymb}       
\usepackage{cite}                    
\DeclareGraphicsExtensions{.pdf}
          
\begin{document}

\title{Consensus Emerging from the Bottom-up: the Role of Cognitive Variables in Opinion Dynamics}

\author{Francesca Giardini$^1$, Daniele Vilone$^{1,*}$, Rosaria Conte$^1$}
\address{$^1$ Laboratory of Agent Based Social Simulation, Institute of Cognitive Science and Technology, National Research Council (CNR), Via Palestro 32, 00185 Rome, Italy.}
\ 

\address{$^*$ Correspondence and requests for materials should be
addressed to D.V.~(email: daniele.vilone@gmail.com).}

\begin{abstract}
The study of opinions $-$ e.g., their formation and change, and their effects on our society $-$ by means of theoretical and numerical models has been one of the main goals of sociophysics until now, but it is one of the defining topics addressed by social psychology and complexity science. Despite the flourishing of different models and theories, several key questions still remain unanswered. The aim of this paper is to provide a cognitively grounded computational model of opinions in which they are described as mental representations and defined in terms of distinctive mental features. We also define how these representations change dynamically through different processes, describing the interplay between mental and social dynamics of opinions. We present two versions of the model, one with discrete opinions (voter model-like), and one with continuous ones (Deffuant-like). By means of numerical simulations, we compare the behaviour of our cognitive model with the classical sociophysical models, and we identify interesting differences in the dynamics of consensus for each of the models considered.
\end{abstract}

\ 


\ 

\maketitle

\ 

\section{Introduction}

Opinions represent a large part of human mental representations, and a large part of our everyday social interactions consist in exchanging, evaluating, revising and comparing opinions with our family, friends, acquaintances, or even strangers. Understanding
opinions, describing how they are generated and revised, and how far
opinions travel across social space both as a consequence of social
influence and as one of the main means through which social influence
unfolds, is crucial for grasping a deeper understanding of human social
cognition and behaviours.

Since the beginning of the last century, sociologists and psychologists have
been interested in understanding opinions, focusing in particular on the multiplicity of
dimensions that characterise this phenomenon~\cite{Kel61,Ols93}. Other disciplines have also been interested in
the topic, like sociophysics and complexity
science~\cite{Heg02,Hel10}. In the last decades, the study of social phenomena as opinion formation and dynamics
has become of great interest in physics. Due to similarities between spreading and ordering phenomena, opinion dynamics has been studied from a
mathematical and numerical point of view by means of the tools of statistical
and computational physics~\cite{Gal08,Cas09}. 
In particular, in the physics community, opinions have been so far considered dynamic elements that can be approximated as spin systems or by similar
statistical-mechanical methods~\cite{Sen14}. Moreover, the possibilities
offered by Big Data Science to collect and analyze huge amounts of digital traces humans leave on the web and other media, has made opinion change and consensus achievement exceeds  disciplinary boundaries and become one of our century's grand scientific challenges~\cite{Laz09}.

Nonetheless, a theory explaining micro-foundations of opinion dynamics is still missing. In cognitive terms, opinions are highly dynamical mental representations resulting from the interplay between different mental objects within individuals' minds. Another distinguishing feature is their being easily affected by opinions of other individuals in the same network, as we will show in our model. Without
explaining how opinions are formed and manipulated within the individuals'
minds, it is very difficult to account for the way in which they change as an
effect of social influence. Even more difficult is to predict consensus or to identify mechanisms leading to polarization or to isolation and/or integration of minorities. This means that when explaining the emergence of macro-social phenomena we need to know what happens at
the micro-level, {\it i.e.}, what drives human actions and decisions in order to
understand how individuals' representations and behaviours can give rise to
socially complex phenomena and how those affect agents' actions. 
Defining an opinion in terms of its mental ingredients and specifying how human cognition promotes opinion change or resistance provides not only a more realistic description of the phenomenon of interest, but it represents an attempt to model the bottom-up emergence of opinions' persistence, the effects of contrasting forces on them and the generation of alternative paths of diffusion.

We aim at developing a cognitively grounded model of opinion dynamics that will allow us to answer the following questions: can we identify the distinctive features of opinions, and model them as interacting representations that get influenced by others' mental states? How can heterogeneous agents, endowed with different representations of the external world, come to share a given viewpoint and what consequences this sharing has on individuals' beliefs and their related behaviours?

Aim of this paper is to put forward a new model of opinion dynamics in which opinions are defined by three main features that interact and support opinion revision and change. We test our model through simulation experiments with two different network topologies and we try to identify how changes at the micro-level of agents' cognition can give rise to changes in opinions at the collective level. In order to check for our models' robustness and reliability, we will also compare our results with existing models, in particular the Voter~\cite{Cli73} model and the work of  Deffuant~\cite{Def00}.

\subsection{Opinions in the Social Sciences}

The study of opinions' formation and spreading originated within the field of social science, in which several important contributions were developed. A comprehensive review of the social and psychological literature on the topic is beyond the scope of this work, but in this section we will tackle the background and we will review some of the main theories that social scientists put forward about opinions and opinion change with the aim of situating our inter-disciplinary model in the wider context of the social sciences. 

In general, opinions are treated as synonyms for different mental objects, as beliefs \cite{Fis75}, or more frequently, attitudes \cite{Ols93,Mcg86,Pri92}. In general, opinions and attitudes are used as interchangeable terms referring to a mental object liable to social influence and persuasion \cite{Pet97}. The specificity of attitudes and the fact that they cannot be considered opinions is made clear in Allport's work~\cite{All35}. In this account, attitudes define and shape people's perceptions, whereas opinions result from a more articulated process in which beliefs, goals, intentions and knowledge play a role. An attitude can be regarded as a necessary element in opinion formation, but it is not sufficient. 
Allport \cite{All67} recognizes the difference between attitudes and opinions, but he nonetheless considers the measurement of opinions as one way of identifying the strength and values of personal attitudes. An alternative view contrastss the affective content of attitudes with the more cognitive quality of opinions that involve some kinds of conscious judgments \cite{Fle67}. In general, it is possible to identify two main trends in the relevant literature.

On one hand, researchers focus on attitudes, considered as implicit evaluations that allow to access individuals' positive or negative views about given matters, and that are supposed to be stable across time and relevant enough to predict individuals' behaviors and actions. On the other hand, explicit theories of opinions are more centered on conscious reasoning and judgment, through which individuals are supposed to form and express measurable opinions. 

Another general feature of the social and psychological approach is the preminence given to measuring, rather than conceptualizing opinions. As a result, many studies (for a review, see~\cite{Sch96}) tried to develop reliable and fine-tuned ways to measure people's approaches to general questions, partially abandoning the issue of defining what an opinion is, and focusing on how it should be measured. Quantifying opinions and measuring their diffusion in a given population is important in order to understand, monitor and predict social and political events, but it is not yet an answer to the question: What is an opinion? Answering this question is essential to define good measures, to identify the right tools and also to get insights from the results. If the measurement does not take into account the main features of opinions, then we will never know what we are exactly measuring and how good our predictions can be, based on that measure.

In political science, Crespi \cite{Cre97} considers individual opinions as "judgmental outcomes of an individual's transactions with the surrounding world" (p.19), emphasizing the interplay between what he calls an attitudinal system and the external world characterized by the presence of other agents and different subjective perceptions. Opinions are the outcomes of a judging process but this does not mean that they are necessarily rational or reasoned, although Crespi recognizes that they need to be consistent with the individual's beliefs, values and affective states. It is worth noticing that many contributions are specifically oriented to understand "public opinion" \cite{Lav96}, as the collective result of integration of opinions and attitudes coming from different sources and exposed to different kinds of influencing.

As other authors already pointed out \cite{Mas07}, many models of opinion and social influence do not provide careful definitions of what an opinion is and how it is affected by social influence. This happens to be true also for theories of persuasion, like the social impact theory \cite{Lat81}, a  theory of how social processes operate at the level of the individual at a given point in time. Part of this theory has been developed using computational modeling by Nowak, Szamrej and Latan\'e \cite{Now90}. In their model, individuals change their attitudes as a consequence of other individuals' influence. In parallel with the idea that social influence is proportional to a multiplicative function of the strength, immediacy, and number of sources in a social force field \cite{Lav96,Lat81} suggest that each attitude within a cognitive structure is jointly determined by the strength, immediacy, and number of linked attitudes as individuals seek harmony, balance, or consistency among them. Therefore, a single opinion does not exist in a "vacuum" and its change depends on the interplay of several factors. Their results show that group dynamics are dependent upon initial spatial configurations, as well as from the different parameters, hence opinion change is completely dependent upon interactions happening at the group level.

\subsection{Tackling the sociophysical background}

The first sociophysical approach to opinion dynamics was the voter model (VM).
It was originally conceived ~\cite{Cli73} as a model of competition of species, but it was soon adapted to model
an electoral competition between two candidates~\cite{Hol75}. The VM dynamics is extremely simple: each agent is
endowed with a binary variable which can assume one of two distinct values. In each elementary interaction an agent is randomly selected and assumes the opinion of one of its neghbours (again  randomly selected). Initial conditions
are also generally set at random. This model has been thoroughly considered in sociophysics because it can be solved exactly
in every dimension~\cite{Cas09} and thanks to its simplicity and flexibility has been used as a starting point in the quantitative treatment of social phenomena. 

In general, the VM dynamics can freeze only with consensus, that is, when all the agents come to share the same opinion. In fact, only in one and two regular
dimensions the system is actually driven towards consensus, following a power-law decay in 1D and a logarithmic
one in 2D. In higher dimensions and on many complex networks the system is not able to reach consensus and remains in a disordered metastable state in the thermodynamical limit~\cite{Soo08} (finite size systems
eventually reach consensus thanks to statistical fluctuations,
after a freezing time which diverges rapidly with size). 

Interestingly, on small-world networks, the level of cooperation of the
metastable state is proportional to the density of long-range connections among distant parts of the system~\cite{Cas03,Vil04}.

However, its simplicity is also one of the main limitations of the model. Attempts to overcome its limits included the application of the original VM in more realistic configurations, meaning that the VM was tested in different kinds of complex topologies~\cite{Cas03,Sla03,Suc05}. Another interesting direction  consisted in adding 
agents with special features with the aim of explaining fundamentalisms~\cite{Mob03} or the effects of mass media~\cite{Gon05}.

The presence of fundamentalists or ``zealots''  who never change opinion, in general does not enhance
consensus: actually, a zealot forces the whole system towards its opinion in one and two regular dimensions, where
consensus is anyway achieved. On the contrary, in different topologies we observe the presence of a small region of localized consensus only in the neighborhood of the fundamentalist~\cite{Mob03,Mob05,Mob07}.

In the VM interactions are based on imitation, a drastic simplification of human interactions. In order to make interaction more realistic, ~\cite{Gal02} proposed a majority rule (MR): at each elementary time step a group of individuals is picked up at random, then all the members
adopt the opinion of the majority of the group itself. This rule was proposed to describe public debates and interactions among individuals belonging to different groups. Also with the majority rule, the system can end up to consensus or to a metastable disordered
state, depending on the details of the topology and the initial conditions. 

In socio-physics, a great deal of work focused on the operational definition of opinion itself. For example, it can be allowed to
assume more than two values~\cite{Wu982} or defined as a continuous variable~\cite{Cas09,Def00}. These models present a richer dynamics and the number of possible active and frozen states accessible to the system is much larger,
so that the phenomena observed are in principle more realistic. Among these models, the Deffuant Model (DM)~\cite{Def00} deserves attention. Here, an opinion is defined as a real number ranging from 0 to 1: therefore, it can be somehow considered a generalization of the VM.  A fundamental feature of DM is the threshold in opinion distance for which
only agents whose opinions are less distant than a given threshold can interact and then get even closer opinions. This mechanism
resembles the Axelrod's model of cultural spreading~\cite{Axe97}, where agents are able to interact proportionally
to their actual cultural similarity. This leads to a final state with a set of different clusters, each one consisting of agents sharing
the same opinion. The number of final clusters depends on the initial conditions and on the model's parameters, especially
the distance threshold~\cite{Def00,Wei01}. Subsequently, the DM has been furtherly refined, for example by defining another
variable characterizing the ``affinity'' among agents~\cite{Bag07}, so that opinions and affinities coevolve: if there is a sufficient affinity between agents, their opinions become more similar, and viceversa. In this version of the model, consensus is reached
more easily than in the original DM.

Although different in several respects, the models mentioned above share the same feature:  opinions are given from the beginning and evolve exclusively in interaction with other agents and on the basis of the environment, without any internal characterization.
No hint of what happens within individuals' minds, how opinions may emerge and evolve internally, is given: in practice, in sociophysics models we have interacting opinions rather than interacting individuals. 

Aim of this work is to overcome the limitations of the social sciences, in which no operational models of opinions are provided, and the limitations of socio-physics just discussed. Moving from a definition of opinions as cognitive representations, we will use socio-physics tools in order to put forward an inter-disciplinary model of opinions which takes into account the internal dynamics of individuals' minds and can be used to analyze opinion spreading and consensus formation. 

\subsection{A tripartite model of opinion: truth-value, confidence and sharedness}

In this work we want to propose an integrated account in which we identify not only the specific cognitive features that characterise opinions, but also the way in which each single feature gets modified by interacting with others' opinions.
A cognitive model of opinions and their dynamics requires to provide a definition of opinions as mental
representations and to present specific features that make their revision and update easy and enduring. 
Moreover, grounding opinions in the mind allows us to take into account not only direct processes of revision due to the interactions with others (social influence), but also revisions based upon changes in one's own mental representations supporting that opinion. 

On the basis of the previous considerations, opinions are mental representations characterised by the following features:

\begin{itemize}
\item subjective truth-value
\item confidence
\item perceived sharedness.
\end{itemize}

We assume that objective truth value cannot be verified (or it is not relevant), so we refer to a "subjective truth-value" that expresses whether and how much someone believes an opinion to be true. Opinions are epistemic representations that refer to matters or events that cannot be defined as either true or false because it is not possible to verify alternative states of the world~\cite{Gia11}, as for example the consequences of an event that did not happen, or the outcomes of an election which has not been held yet. In many cases opinions are emotionally loaded, and individuals have positive or negative attitudes that reinforce their opinions, as for instance about abortion~\cite{Cre97}. In that case there may be not any ascertainable truth about it, but people often have strong opinions on the basis of their beliefs, culture or personal experience. In order to convert a subjective truth-value into a practical quantity, we will express it as a variable indicating how true an opinion is considered by a given agent. 

The second defining feature of our model is the degree of confidence. This is a subjective measure of the strength of an opinion and it expresses the exent to which one's opinion is resistant to change, like a sort of ``opinion inertia''. When an individual is highly confident, he has a number of reasons to believe that his/her opinion is right, and the higher the confidence, the more willing that individual is to defend his/her opinion against others'. For the sake of simplicity, confidence is currently expressed as an arbitrarily assigned value. It is worth stressing the difference between subjective truth-value and degree of confidence: an individual can be somehow uncertain about his/her opinion, but struggle fiercely for it (``right or wrong, it's my country''); or, viceversa, he can be absolutely certain about a given issue without defending it publicly (for fear or lack of interest).

The third feature of our model is perceived sharedness, {\it i.e.}, the extent to which a given opinion is perceived to be shared. This subjective measure allows us to distinguish between the actual number of individuals having the same opinion and the personal experience that an agent can have. Believing that there are a number of other individuals sharing my opinion does not mean that this is actually the case, actually I can be part of a minority and not be aware of the opinion of the majority. Perceived sharedness may heavily affect degree of confidence, making people feel more confident because they might feel supported by the majority.  On the other hand, this factor can also lead an individual to revise his/her opinion because confidence in it is low and it is also perceived as a minority opinion. 

Modeling opinions as results of the interplay among the above defined internal variables allows us to take into account their distinctive nature of mental objects. Knowing how opinions change within individuals' minds is essential to understand whether this change can be robust, how far it can travel into a given social group and how far it can spread. Unlike other models that treat opinion change as an effect of plain social influence~\cite{Now90}, we are interested in modeling what happens inside individuals' minds. Opinions do not change suddenly, but they result from interacting internal dynamics and external influences, as our model is aimed to show. 

\ 

\section{Discrete Opinion Model}

\subsection{Description}

We set $N$ individuals which are characterised, at every time step $t$, by their opinions $O_i(t)$ ($i=1,\dots,N$).
In the discrete version of the model, opinions can
assume one of two possible values, 0 or 1. Agents
are characterised by three internal variables referring to their opinions: subjective truth-value $x_1^i(t)$,
degree of confidence $x_2^i(t)$, and perceived sharedness $x_3^i(t)$. These variables are defined as continuous
quantities, ranging from 0 to 1. 

Agents are located on the nodes of a
graph, with the relationships among them given by the links. At the beginning of the simulation,
agents are initialized as having one of two opinions which are randomly assigned with equal probability (with some exceptions which will be pointed out when needed). Internal variables are assigned at random as well, following a uniform distribution.
For each elementary time step, two agents are randomly chosen and they are attributed one of two roles: a listener, say $i$, and a speaker, $j$.
Speakers are selected among listener's neighbours. In each interaction, evolution applies only to listeners, which, in our asymmetric model, are those
who can revise their opinions. It is worth noticing that in our model, even though two agents share the same opinion, this can result from completely different internal variables. In particular, opinions and internal variables coevolve. People sharing the same opinion reinforce their beliefs when they meet, conversely interacting with an agent of opposite opinion
may drive individuals to change it or at least to be less sure of their initial opinion. 

In our model, the subjective truth-value
and the degree of confidence of a listener increase (or remain unchanged) when interacting with a speaker sharing the
same opinion. In formal terms, we implement the dynamics of these internal variables when $O_i(t)=O_j(t)$ as follows

\begin{enumerate}
  \item If $x_1^i(t)<x_1^j(t)$ and $x_2^i(t)<x_2^j(t) \Longrightarrow x_1^i(t+\Delta t)=x_1^j(t)$;
  \item If $x_1^i(t)\geq x_1^j(t) \Longrightarrow x_1^i(t+\Delta t)=x_1^i(t)$;
  \item If $x_2^i(t)\geq x_2^j(t) \Longrightarrow x_2^i(t+\Delta t)=x_2^i(t)$;
  \item If $x_2^i(t)<x_2^j(t) \Longrightarrow x_2^i(t+\Delta t)=\frac{x_2^i(t)+2x_2^j(t)}{3}$.
\end{enumerate}

In practice, when a listener is paired with a speaker sharing the same opinion, its subjective truth-value
cannot decrease. In particular, we assume that if the speaker is more confident than the listener, the latter is reinforced in
his/her opinion; vice versa, a speaker with a low degree of confidence can marginally influence a listener. We implemented this rule in the simplest way:
the listener will assume the speaker's subjective truth-value ($x_1^j$) if and only if both confidence and subjective truth-value are larger in the speaker,
otherwise $x_1^i$ does not change.
Rules 3 and 4 define how confidence changes as an effect of interactions. If a listener meets a more confident
speaker with the same opinion, this will increase the listener's confidence, otherwise it will not be affected.
Unlike subjective truth-value, the listener does not assume the speaker's confidence, but there is an adjustment
between the two values. 

Let us now define the rules for the dynamics of the opinion and the internal variables when a speaker $j$ and a listener $i$ of different opinions meet.
In this case, opinion change depends on confidence and perceived sharedness. 
In general, we assume that people can be better persuaded by more confident speakers, in line with evidence in social psychology showing that consistent and confident minorities can influence majorities and make them revise their judgements~\cite{Mos76}.
We also consider that there is a positive correlation between the speaker's subjective truth value and the probability of convincing the listener. Finally, humans are sensitive to social pressure, then believing that one's own opinion is shared by the majority makes opinion change less likely~\cite{Kel58}. 
Therefore, we define a simplified rule such that the probability $P_{ij}$
that $i$ assumes the $j$'s opinion is zero if speaker's confidence is smaller than the listener's, otherwise it is directly
proportional to the speaker's subjective truth-value and to the quantity 1 minus the perceived sharedness. This points to the fact that
an agent is reluctant to change its opinion if thinking to share it with the majority. Summarizing, it is

\begin{equation}
P_{ij} =
\left\{
\begin{array}{ccc}
x_1^j(t)[1-x_3^i(t)] & \mbox{if} & x_2^j > x_2^i \\
 & \\
0 & \mbox{if} & x_2^j\leq x_2^i \ ;
\end{array}
\right.
\label{pij}
\end{equation}

Afterwards, if the listener shifts towards the speaker's opinion, we assume that the former also acquires the subjective truth-value
of the latter, but with a confidence level equaling half of the speaker's confidence.

\begin{equation}
  \begin{array}{l}
x_1^i(t+\Delta t)=x_1^j(t) \\
 \\
x_2^i(t+\Delta t)=x_2^j(t)/2 \ .
\end{array}
\label{int1}
\end{equation}

On the contrary, if $i$ does not change opinion, the listener's subjective truth-value remains unchanged, but
the interaction affects in any case the confidence. More precisely, we assume that interacting with someone
with a different opinion reduces the listener's degree of confidence.
Therefore, we set the new listener's confidence as the average between its own confidence before the interaction
and the speaker's confidence level in case the latter was smaller, to half its original value if initially it
was $x_2^i<x_2^j$.

\begin{equation}
x_1^i(t+\Delta t)=x_1^i(t)
\label{int2_a}
\end{equation}

\begin{equation}
x_2^i(t+\Delta t)=
\left\{
  \begin{array}{cl}
\frac{x_2^i(t)+x_2^j(t)}{2} & \ \ \mbox{if}\ \ \ \ \ x_2^i(t)>x_2^j(t)  \\
 & \\
\frac{x_2^i(t)}{2} & \ \ \mbox{if}\ \ \ \ \ x_2^i(t)\leq x_2^j(t)  \ . 
\end{array}
\right.
\label{int2}
\end{equation}

Finally, we implemented perceived sharedness $x_3^i(t)$ as the total number of past encounters with other agents sharing the same
opinion an agent $i$ has up to time $t$, independently of the outcomes of each interaction. More precisely, each agent
records all past encounters as a listener, determining the frequency of both opinions: the perceived sharedness at
time $t$ is the frequency of the opinion shared by the agent itself at the same instant $t$.

Time is measured in Monte Carlo steps. This means that a single time unit is made up by $N$ single interactions,
{\it i.e.} $\Delta t\equiv1/N$.

Certainly, this model must be seen as a first step, and further studies and in-depth analyses will be useful for a more precise
definition of its parameters.

\subsection{Topology}
We tested the behaviour of our model on two different topologies: a total connected graph
(mean field) and a one-dimensional ring. We chose them in accordance with the relevant literature~\cite{Cas09}, in which these topologies are usually
considered as starting points. Moreover, given the complexity of the relationships among agents' internal
variables, even simple topologies can be useful in understanding the basical properties of the model.
More complex topologies will be used in future works.

\subsection{Results}

For each version of the model, we initially considered a system
of $N=100$ agents and tested its behaviour starting from three different initial distributions of agents holding two different opinions: $\langle O_0\rangle=0.5,\ 0.75$ and $0.9$. The last two distributions represent an attempt to model a situation in which there is a conflict between a majoritarian and a minoritarian opinion. 
We stress the fact that every measure represents an average over 2000 independent realizations.
More details are given in the following subsections.

\subsubsection{Mean field}

In every simulation the system ends up to a substantial consensus. At the end of each realization, all the agents share the same opinion, even if there are interesting differences between individuals' mental states.

\begin{figure}[h]
\begin{center}
\vspace*{1cm}
\includegraphics[width=10cm]{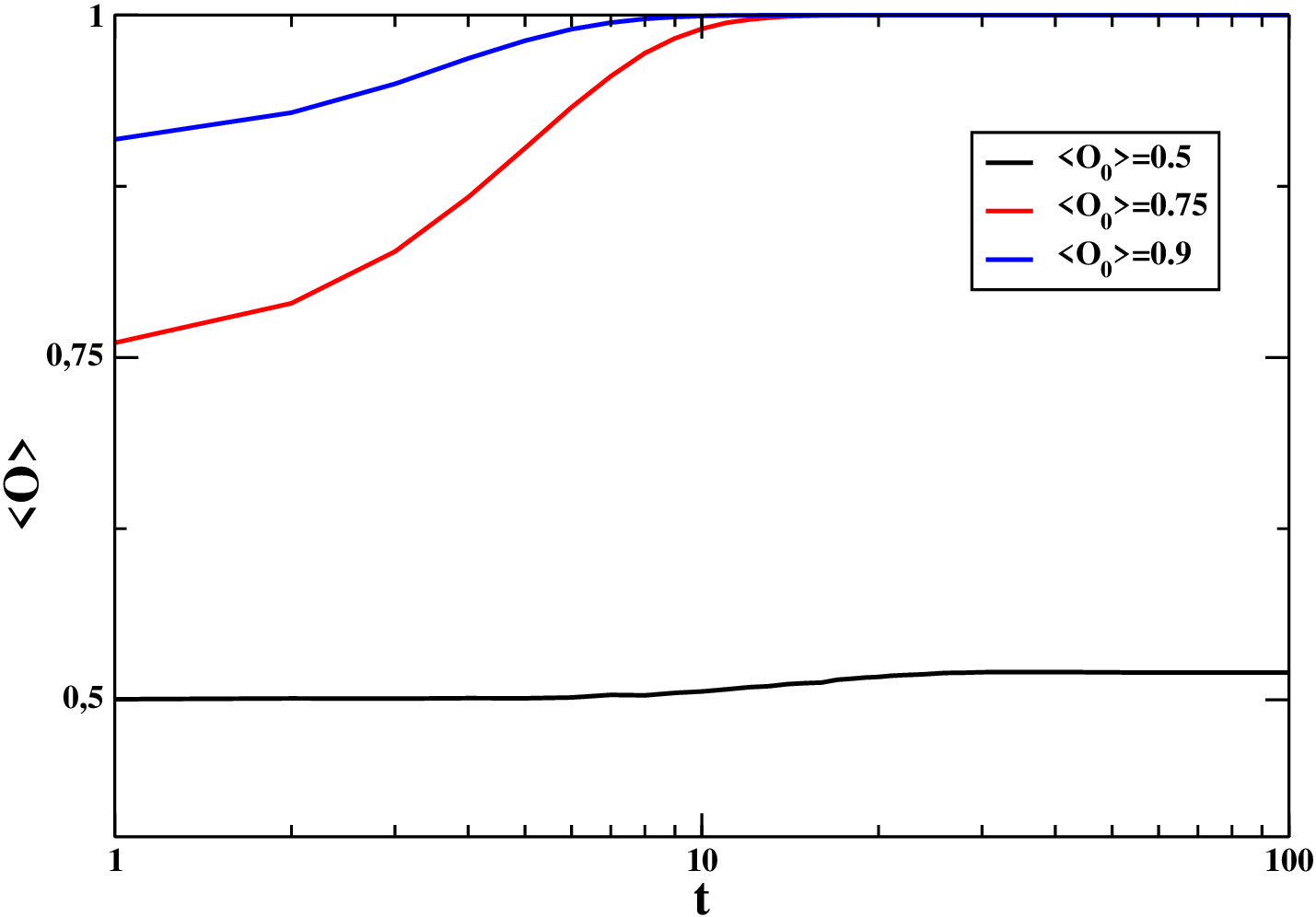}
\end{center}
\textbf{\refstepcounter{figure}\label{discr_Ot} Figure \arabic{figure}.}{ Average opinion as a function of time for a system of $N=100$ on a
totally connected graph for three different initial conditions. Figure presents values averaged over 2000 independent realizations.}
\end{figure}

\begin{figure}[h]
\begin{center}
\vspace*{1cm}
\includegraphics[width=10cm]{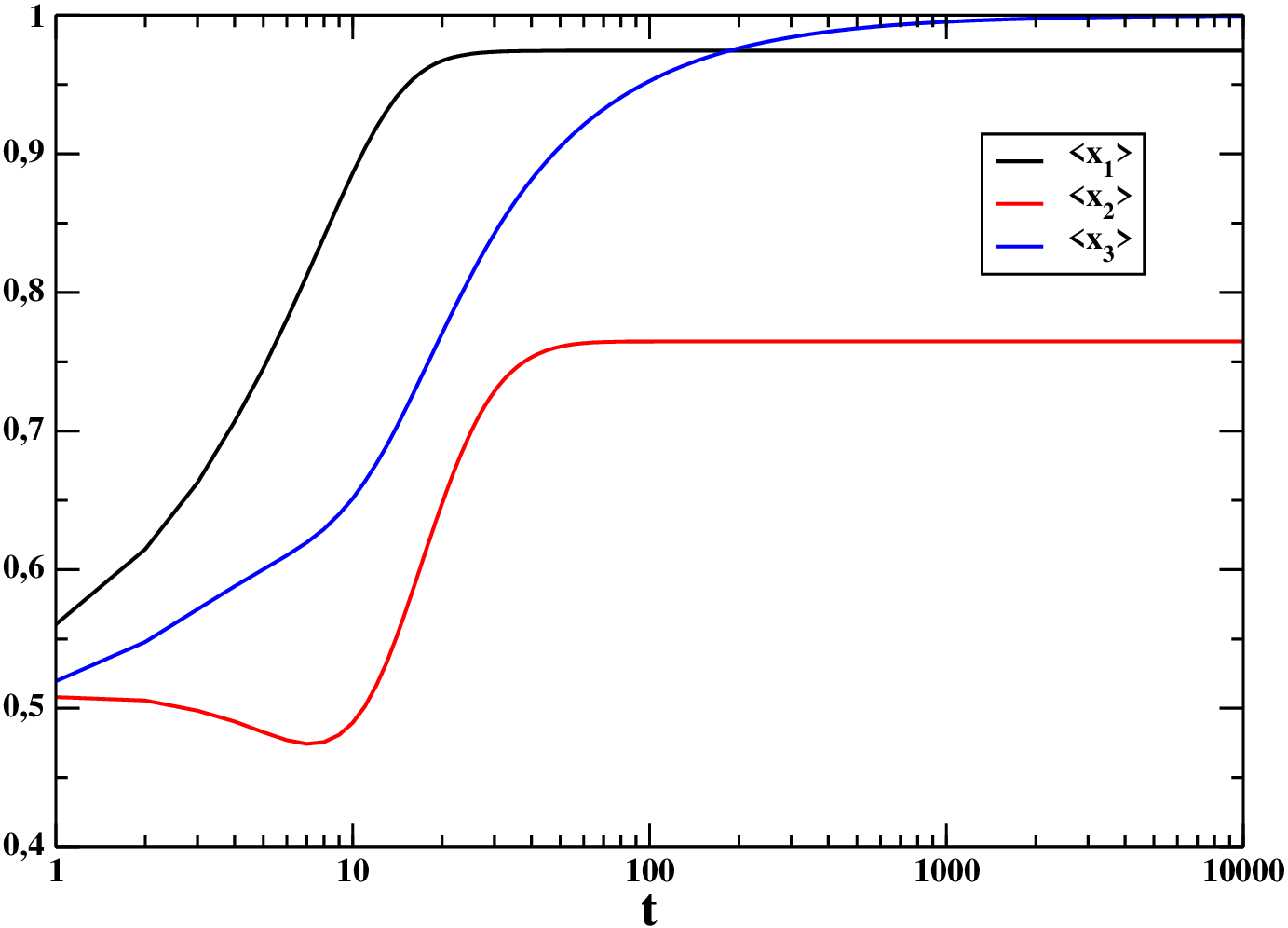}
\end{center}
\textbf{\refstepcounter{figure}\label{discr_Intv} Figure \arabic{figure}.}{ Time behaviour of average internal variables for
a system of $N=100$ on a totally connected graph and initial average opinion $\langle O_0\rangle=0.5$.
Averages over 2000 independent realizations.}
\end{figure}

In Figure~\ref{discr_Ot} we show how average opinion behaves over time. It is worth noticing that an initial asymmetry, even if small,
allows the majoritarian opinion to invade the whole system and to become dominant. That means that the opinion which at $t=0$ is majoritarian,
even slightly, always ends up being the only survived one. This result contrasts with the classical mean field voter model, in which the
initial average opinion is always conserved~\cite{Cas09,Sen14}.

On the other hand, in Figure~\ref{discr_Intv} we show the behaviour over time of the averaged internal variables with $\langle O_0\rangle=0.5$. Among these, perceived sharedness shows a noticeable behaviour: it takes longer to reach its final level (which equals 1 since all the agents share the same opinion eventually).
In other words, the system reaches consensus even before agents realize they all agree, showing an interesting dynamics between the
micro- and the macro-level. This result is remarkable especially if we think of real-world situations, in which there is usually a
gap between what individuals know locally and the global and emergent behaviour of the system in which they are embedded. This delay
is due to the early encounters at the initial stages of the dynamics, kept in memory by the agents, which drive the individuals to
mantain a distorted idea of the others' views for a longer time. 
It is worth to notice that the persistence of the initial opinion and impressions in human mind is a well-known phenomenon, already
discussed in the psychological literature~\cite{Dav97,Woo14}.

In order to be sure that the observed results do not depend on the size of the system, we also repeated the same simulations with $N=250$: we did not
observe any significant differences increasing the number of agents in the population.

It must be noticed that even if this dynamics leads the system to consensus, in half of the realizations there is one agent who maintains its minoritarian opinion. This resistance to social influence happens when the agent with the
largest initial degree of confidence has the opposite opinion of the majority: from
Equation~(\ref{pij}) follows that such agent can never change opinion. Such a result could help explaining extremisms in real-world scenarios, where no matter the kind of social influence and persuasion someone is exposed to, there are always individuals who stick to their opinions. 

\subsubsection{One-dimensional ring}

We tested our model also on a one-dimensional ring with the same number of agents ($N=100$). 

\begin{figure}[h]
\begin{center}
\vspace*{1cm}
\includegraphics[width=10cm]{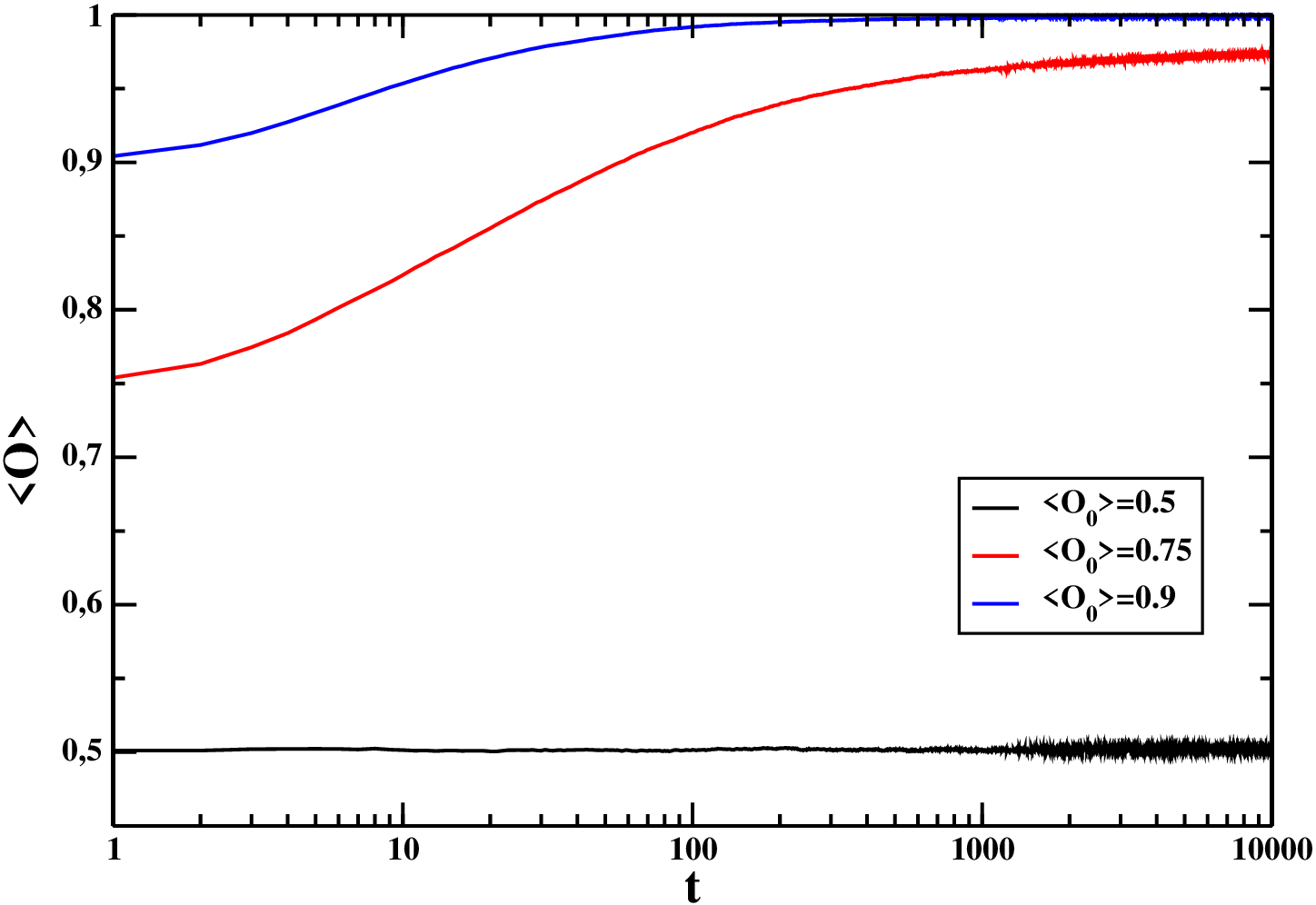}
\end{center}
\textbf{\refstepcounter{figure}\label{1D_Ot} Figure \arabic{figure}.}{ Average opinion as a function of time for a system of $N=100$ on a
one-dimensional ring for three different initial conditions. Averages over 2000 independent realizations.}
\end{figure}

\begin{figure}[h]
\begin{center}
\vspace*{1cm}
\includegraphics[width=10cm]{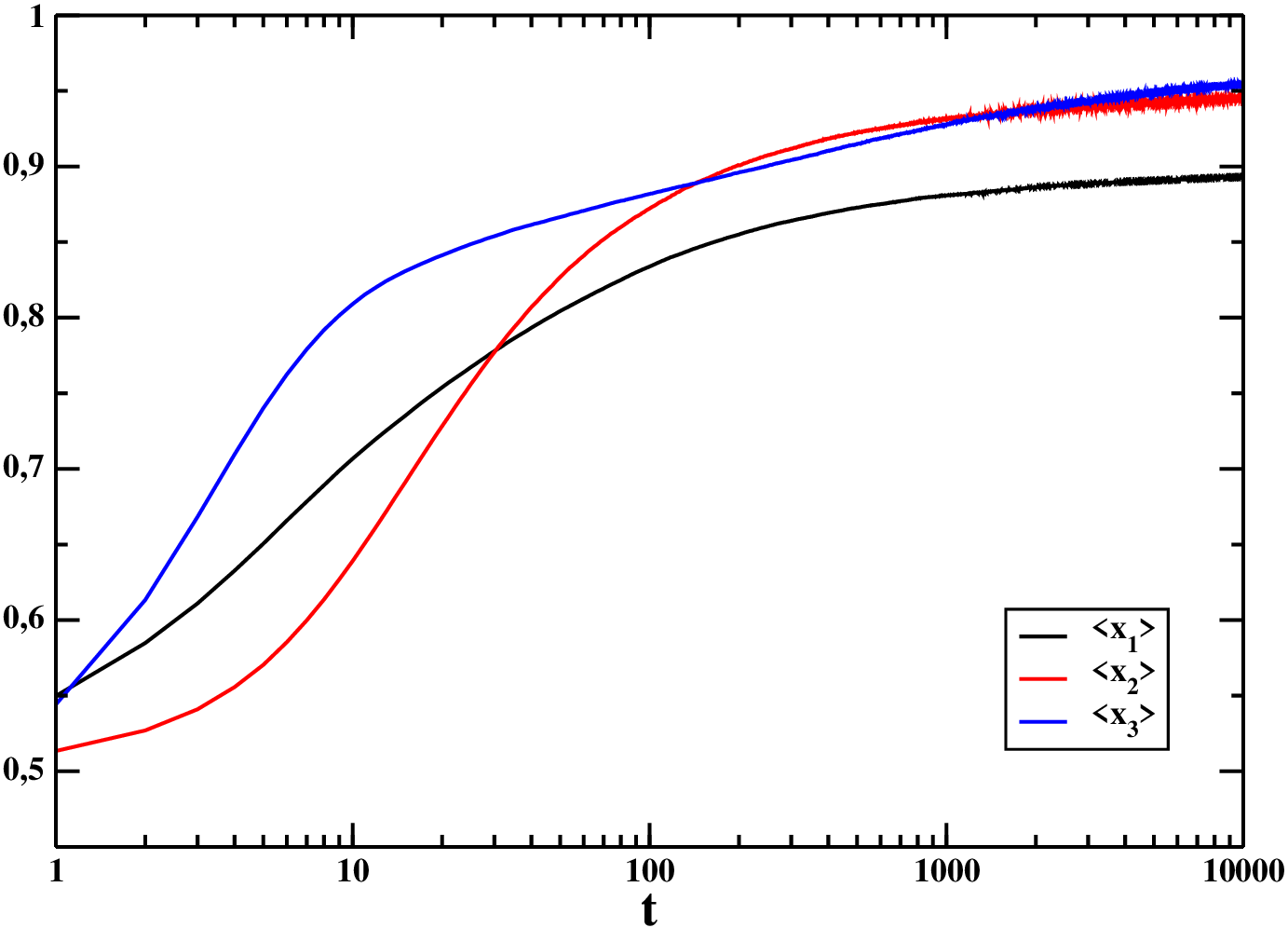}
\end{center}
\textbf{\refstepcounter{figure}\label{1D_Intv} Figure \arabic{figure}.}{ Behaviour over time of average internal variables for
a system of $N=100$ on a one-dimensional ring and initial average opinion $\langle O_0\rangle=0.5$. Averages over 2000 independent realizations.}
\end{figure}

\begin{figure}[h]
\begin{center}
\vspace*{1cm}
\includegraphics[width=10cm]{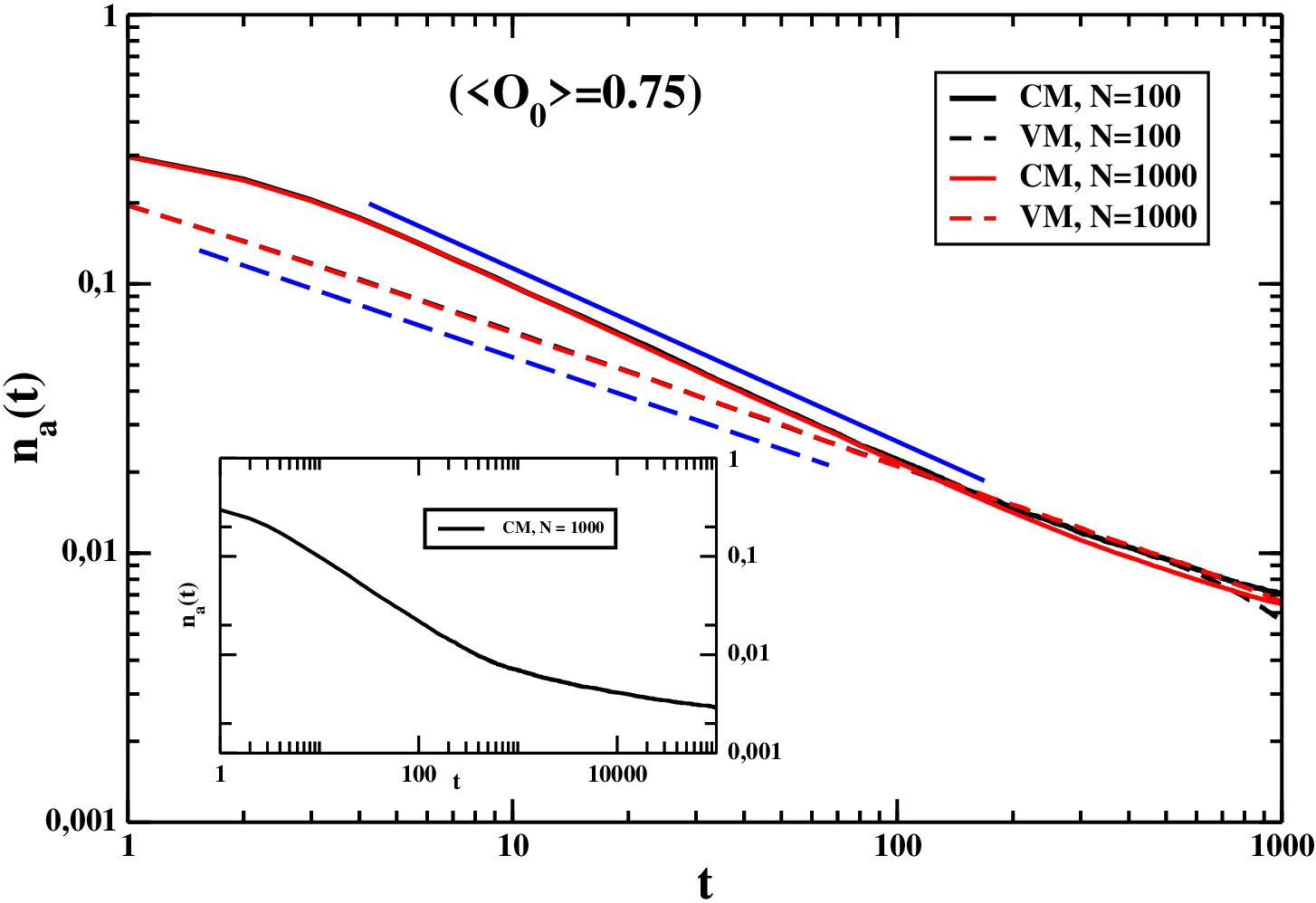}
\end{center}
\textbf{\refstepcounter{figure}\label{paredes} Figure \arabic{figure}.}{ Behaviour over time of the (average) active bond density
for the cognitive opinion model (CM) on a one-dimensional ring and initial average opinion $\langle O_0\rangle=0.75$.
System size: $N=100$ (full black line) and $N=1000$ (full red line).
Comparison with the voter model with the same parameters (dashed lines: black for $N=100$, red for $N=1000$).
The blue dashed line represents a power decay with exponent $1/2$, the blue straight one a power decay with exponent $2/3$.
{\bf Inset:} time behaviour of the active bond density for CM with $N=100$ up to $t=10^5$. Averages over 2000 realizations.}
\end{figure}

Figures~\ref{1D_Ot} and~\ref{1D_Intv} show how average opinion (for three different initial conditions) and internal
variables (for $\langle O_0\rangle=0.5$) change over time, respectively. Concerning the average opinion, qualitatively
these results are similar to the mean field case, even though here the invasion of the initially majoritarian opinion
is much slower. This behaviour is confirmed in Figure~\ref{paredes}, where we compare the time behaviour
of active bonds ({\it i.e.} the links between two agents with opposite opinion) in our model and in the one-dimensional
voter model~\cite{Cas09,Vil04}, in which, at each time step, a randomly selected agent simply imitates the opinion of one of its neighbours.
This makes the system reach consensus through a power-law decay with exponent $1/2$
(and a final quick exponential convergence time depending on the square of the system size),
whereas a different behaviour is observed in our model. After an initial power-law decay with exponent $\beta$ close to  $2/3$ (a numerical
fit reports $\beta=0.65\pm0.01$),
the active bond density decreases much more slowly (maybe tending to a steady state), as shown in the inset of the same figure.

\ 

\section{Continuous Opinion Model}

\subsection{Description}

We also present another version of the model in which opinions are continuous, so that an opinion is a real variable ranging between 0 and 1
(initially randomly assigned to each agent following a uniform probability distribution).
The model dynamics is the same we described above, but we extended its scope by adding a new rule for opinion revision. Here, when
the listener accepts the speaker's opinion, opinion changes in accordance with Deffuant's rule~\cite{Def00,Wei01}:

\begin{equation}
O_i(t+1)=O_i(t)+\mu[O_j(t)-O_i(t)] \ ;
\label{contin}
\end{equation}

with $\mu=0.5$. In order to determine the perceived sharedness, we consider opinions up to 0.5 as of ``negative'' kind
and from 0.5 as ``positive'', so that $x_3^i(t)$ is the number of past encounters with other agents sharing the same
kind of opinion an agent $i$ has up to time $t$, starting from $t=0$.
The internal variables are assigned at random, following a uniform distribution, also in this version of the model.

\subsection{Results}

\subsubsection*{Mean field}

\begin{figure}[h]
\begin{center}
\vspace*{1cm}
\includegraphics[width=10cm]{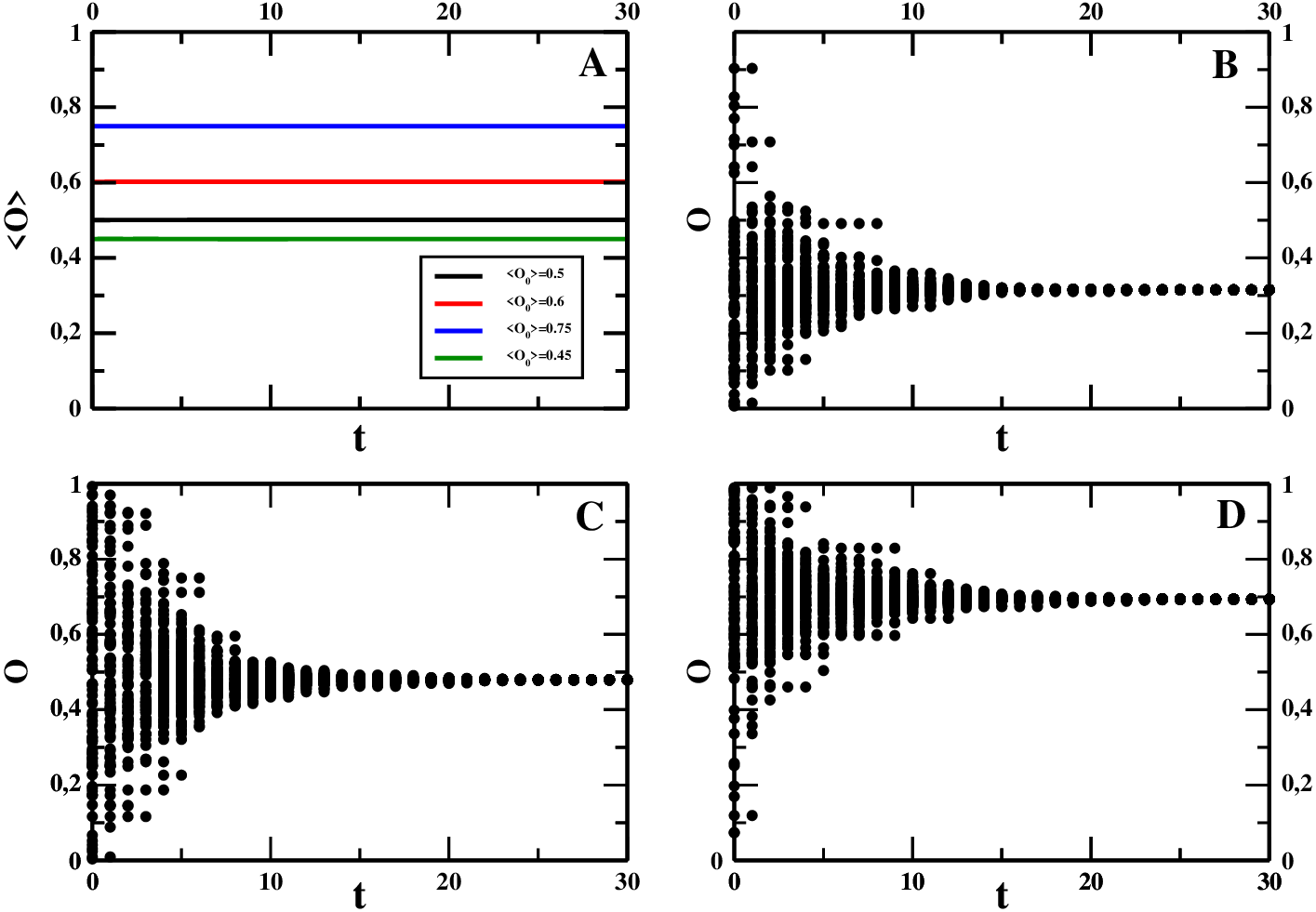}
\end{center}
\textbf{\refstepcounter{figure}\label{cont_Ot} Figure \arabic{figure}.}{ Behaviour of the opinion for a system of $N=100$
individuals on a totally connected graph; continuous opinion. Averages over 2000 independent realizations.
{\bf A}: Time evolution of the average opinion for four different initial conditions (from top to bottom of the figure:
$\langle O_0\rangle=0.75,\ 0.6,\ 0.5,\ 0.45$, respectively), average over 2000 independent realizations.
{\bf B}, {\bf C}, {\bf D}: Evolution of the opinion of each agent during a single realization, for initial average opinion $\langle O_0\rangle=0.3$,
$0.5$ and $0.7$, respectively.
 }
\end{figure}

\begin{figure}[h]
\begin{center}
\vspace*{1cm}
\includegraphics[width=10cm]{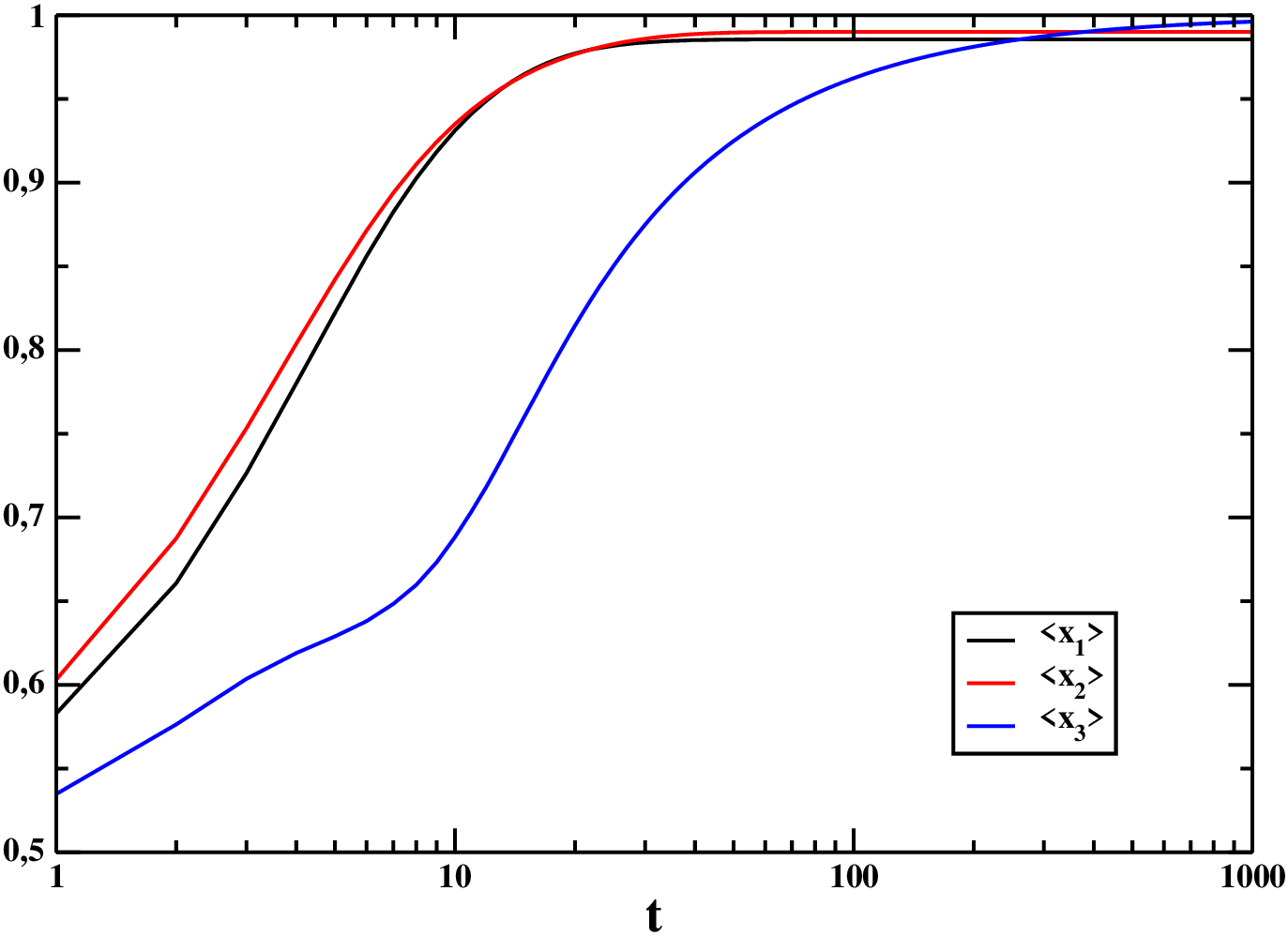}
\end{center}
\textbf{\refstepcounter{figure}\label{cont_intv} Figure \arabic{figure}.}{ Time behaviour of the average internal variables for
a system of $N=100$ on a totally connected graph and initial average opinion $\langle O_0\rangle=0.5$; continuous opinion.
Averages over 2000 independent realizations.}
\end{figure}

In the continuous version of the model, when the graph is totally connected, the
system reaches final consensus, even though conserving the average initial opinion, as we can see in Figure~\ref{cont_Ot}.
On the other hand, internal variables behave as in the discrete case, as shown in Figure~\ref{cont_intv}. This means that when we apply a mean field topology, the model shows the same behaviour, regardless of the way in which opinions are implemented (either as discrete or as continuous variable): the system achieves final consensus with high
levels of subjective truth-value and degree of confidence, reached even before the agents realize they have already ended up sharing the same opinion.

\ 

\section{Discussion}

Opinions are a complex issue for a variety of reasons and they represent a very
interesting case of the micro-macro link: opinions are mental objects that get
modified by social processes and then re-enter the mental space. More specifically,
internal representations of individuals, such as beliefs, goals, and intentions, give
rise to the complex dynamics at the macro-level, which feed backs into the lower
level. However, pivotal to understanding the dynamics of opinions is the definition of specific features, which characterise how opinions are created and modified within individual minds. 

In this work, we have outlined the micro-level of opinions
and started to explore their dynamics. We identified three main constitutive features of
opinions within individuals' minds. The first internal variable
is the subjective truth-value, which describes how much the individual believes her/his
opinion to be true. Saying that opinions have a subjective truth-value does not mean that
individuals do not believe in them or that they cannot be strongly
opinionated. This feature is crucial because it accounts for the fact that
opinions get more frequently influenced by interactions with others through
social influence, but they can also be easily revised according to changes in
one's own mental representations, without any external influence. 
The second feature is what we called degree of confidence, which defines to what extent an individual trusts his/her opinion and how much the opinion is resistant to
change. 
The last variable we introduced is perceived sharedness, {\it i.e.}, the extent to which one
believes others share the same opinion, thus providing support for it.
Perceived sharedness does not necessarily overlap with the
actual sharedness, and there can be a gap between how much someone
believes an opinion to be shared and the actual number of individuals
sharing it. 

Our results show that the interaction of these variables at the micro-level generates interesting macro-dynamics. Compared with the voter model, we observe interesting differences. The voter model dynamics has two main features: the magnetization is conserved, but in a mean field topology and one dimension, consensus is reached for every finite system size~\cite{Cas09,Sen14}. This means that,
even though in each realization all agents end up sharing the same opinion
(0 or 1), the average over the ensemble always conserves the initial average opinion. In our discrete model, instead, as shown in Figure~\ref{discr_Ot}, this is true only for totally
random initial configurations ({\it i.e.}, $\langle O_0\rangle=0.5$). On the contrary, if simulations start with an even small asymmetry between the two opinions, the most spread one outcompetes the other. 

\begin{figure}[h]
\begin{center}
\vspace*{1cm}
\includegraphics[width=10cm]{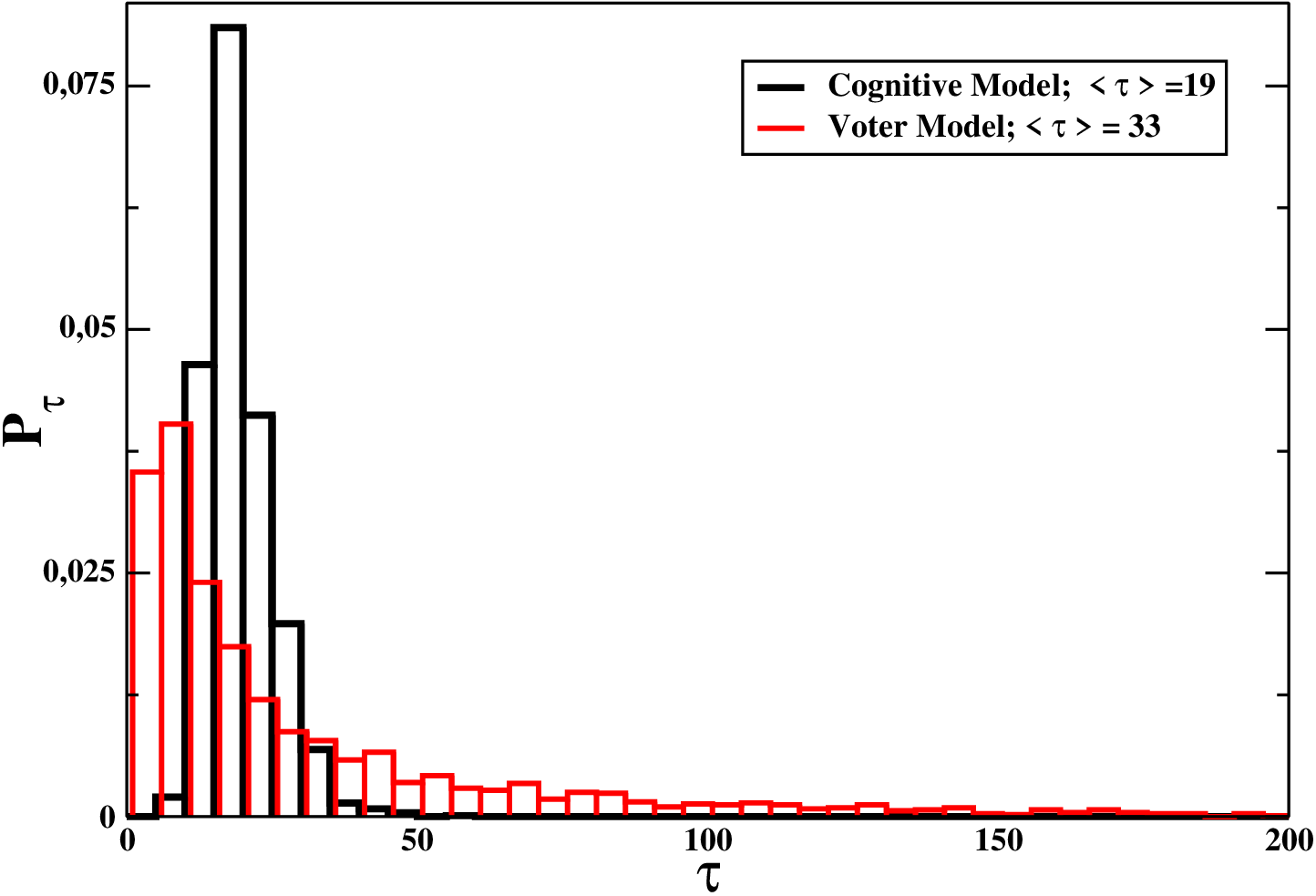}
\end{center}
\textbf{\refstepcounter{figure}\label{tau_distr} Figure \arabic{figure}.}{ Probability distribution of the time needed to reach consensus $\tau$
for the cognitive opinion model (black) and the voter model (red). Mean-field case, system size $N=100$, initial average opinion
$\langle O_0\rangle=0.5$.}
\end{figure}

Another difference between the voter dynamics and our cognitive model consists in the amount of time needed to reach the final totally ordered state, which strongly depends on the system topology.
On a totally connected graph our model is faster than VM. Actually,
even though the convergence to the final configuration is in both cases exponential,
the distribution of consensus times $\tau$, {\it i.e.} the time needed to reach the consensus,
is rather different, as shown in Figure~(\ref{tau_distr}). Both distributions
have indeed the same mode and an exponential tail, but due to its longer tail, 
the average $\tau$ for the voter model is almost two times larger. Moreover, the latter has a much larger
standard deviation ($\sigma_{\tau}\simeq44$ for the voter model, and ($\sigma_{\tau}\simeq5$ for the cognitive one),
meaning that also consensus times very far away from the average are likely to be observed.

Conversely, in one dimension we observe that our cognitive model is slower than the voter one. Actually,
as it is easy to understand from Figure~\ref{paredes}, the active bond density starts decaying similarly to the voter model
(even though with a different power-law exponent), but after a certain time
$t^*$ it reaches a sort of plateau whose level decreases slightly on the system size. In short,
for our model the time behaviour of active links can be written as follows:

\begin{equation}
n_a(t) \propto
\left\{
\begin{array}{lcc}
t^{-\beta} & \mbox{if} & t\lesssim t^* \\
 & \\
g(t) & \mbox{if} & \ \ t\gtrsim t^* \ ,
\end{array}
\right.
\label{scaling2}
\end{equation}

where $g(t)$ is a decreasing function of time.
From Figure~\ref{paredes}, it is possible to infer that the decay exponent is $\beta\simeq0.65$.
On the other hand, the behaviour of $g(t)$ and the value of $t^*$, which {\it a priori} could also depend on $N$,
are very hard to be assessed precisely. Their determination would require the numerical study of much
larger systems up to very large times, averaged over more than thousands of realizations,
but such effort would go beyond the scope of this work.
Anyway, while an exact form of the Equation~(\ref{scaling2}) may be obtained only through further investigations, the main trend
(that is, a slow convergence to the consensus with cognitive dynamics of opinions) is already clear.

In mean field topology our model shows that final consensus is achieved faster, and this is true even when taking into consideration continuous opinions. Actually, comparing Figure~\ref{cont_intv} with the results for
similar systems reported in~\cite{Wei01}, it can be easily noticed that
the latter needs much more time to reach the final ordered state, if compared with our results.

Our work represents a first attempt to merge two related but usually distant approaches. Cognitive modeling allowed us to define the basic elements of an opinion as a mental representation characterised by specific features, whereas we used the tools of socio-physics to model the dynamics of opinion spreading. Further investigation is needed in order to better understand opinion change and to set more properly the parameters at stake, but also to predict whether these changes will be stable and under what conditions. 
In particular, we need to investigate more exstensively the interplay among different dynamics. Opinions change in the mind
possibly under social influence, and the social cognitive dynamics of opinions affects the way they spread and the configuration
of the space of opinions. 

On the other hand, the initial topology may bear consequences on opinion change and on mental
features that characterise existing opinions. Furthermore, the interplay between opinions and other mental objects, like
other types of beliefs, ought to be addressed, and the external validation of the cognitive model against real-world data
should be also achieved. Finally, a well-determined comparison between cognitive and classical socio-physical opinion models,
here outlined, would be useful to understand more correctly merits and limits of both approaches.

\section*{Disclosure/Conflict-of-Interest Statement}

The authors declare that the research was conducted in the absence of any commercial or financial relationships that could be construed as a potential conflict of interest.

\section*{Author Contributions}

Francesca Giardini and Rosaria Conte developed the model, Daniele Vilone performed numerical simulations and analyzed data, and all the authors wrote the paper.

\section*{Acknowledgments}

This material is based upon work supported by CLARA (CLoud plAtform and smart underground imaging for natural Risk Assessment), funded by the Italian Ministry of Education and Research (PON 2007-2013: Smart Cities and Communities and Social Innovation; Asse e Obiettivo: Asse II - Azione Integrata per la Societ\`a dell'Informazione; Ambito: Sicurezza del territorio).

\section*{References}

\bibliography{GVC15_arXiv_revised}{}
\bibliographystyle{unsrt} 

\end{document}